# MODIFIED LLL ALGORITHM WITH SHIFTED START COLUMN FOR COMPLEXITY REDUCTION

Nizar OUNI[1] and Ridha BOUALLEGUE[2]

[1]National Engineering School of Tunis, SUP'COM, InnovCom laboratory, Tunisia
[2]SUP'COM, InnovCom laboratory, Tunisia

## ABSTRACT

*Multiple-input multiple-output (MIMO) systems are playing an important role in the recent wireless communication. The complexity of the different systems models challenge different researches to get a good complexity to performance balance. Lattices Reduction Techniques and Lenstra-Lenstra-Lovàsz (LLL) algorithm bring more resources to investigate and can contribute to the complexity reduction purposes.*

*In this paper, we are looking to modify the LLL algorithm to reduce the computation operations by exploiting the structure of the upper triangular matrix without "big" performance degradation. Basically, the first columns of the upper triangular matrix contain many zeroes, so the algorithm will perform several operations with very limited income. We are presenting a performance and complexity study and our proposal show that we can gain in term of complexity while the performance results remains almost the same.*

## 1. INTRODUCTION

MIMO communication systems are used to provide high data rate. Basically, the MIMO system consists of transmitting multiple independent data symbols over multiple antennas. At the receiver side, a MIMO decoder need to be used to detect, separate, and reconstruct the received symbols. Several linear detection schemes can be used, such as the zero-forcing (ZF) or the minimum mean square error (MMSE) criterion, Maximum likelihood (ML) is consider as the optimal solution for the MIMO detection. But, unfortunately the ML algorithm remains complex for hardware implementations. Therefore, linear MIMO detection techniques like ZF and MMSE seems to be suitable in term of complexity, but suffers from bit error-rate (BER) performance degradation.

During last years, a lattice-reduction (LR) pre-processing techniques has been proposed to be used with linear detection in order to transform the system model into an equivalent system with better effect channel matrix.

The populated LR algorithm is called the Lenstra-Lenstra-Lovàsz (LLL) algorithm is the most used one. It was called according to the name of the inventors [1]. But, the LLL algorithm brings many challenges due to higher processing complexity and the nondeterministic execution time [2].

Multiple other varieties of the LLL are presented, such as [4] and [5] where the goal was to give a good complexity to performance balance.

In this paper, we will focus on the ZF decoding technique and we propose a modification to the original LLL algorithm to reduce the number of loops by shifting the iteration start point. This reduces the complexity of the algorithm and keeps the BER degradation negligible.

## 2. SYSTEM MODEL DESCRIPTION

During this paper we will consider that $(.)^H$ and $(.)^T$ denote respectively the hermitian transpose and the transpose of a matrix.

We consider the spatial multiplexing MIMO system with $N_t$ transmit and $N_r$ receive antennas with a Rayleigh channel non variant in the time.

$$x = H.s + n \qquad (1)$$

Where $s = [s_1, s_2, \dots, s_{N_t}]^H$; $(s_i \in s)$ is the information vector with $S$ being a constellation set of square quadrature amplitude modulation(QAM) with $E[ss^H] = \sigma_s^2.I_{N_t}$ and the real and imaginary parts are $\{-\sqrt{M_S} + 1, \dots, -1, 1, \dots, \sqrt{M_S} - 1\}$ with $M_S$ being the constellation size. We will suppose that the average transmit power of each antenna is normalized to one, so $E[ss^H] = I_{N_t}$. With $I_m$ is the m $\times$ m identity matrix.

$H$ is an $N_r \times N_t$; $(N_r \geq N_t)$ complex channel matrix, $x = [x_1, x_2, \dots, x_r]^T$ is the received signal vector, and $n = [n_1, n_2, \dots, n_{N_r}]^T$ is the complex additive white Gaussian noise (AWGN) vector with zero mean and covariance $\sigma_n^2.I_{N_r}$.

On the receiver side, $x = [x_1, x_2, \dots, x_{N_r}]^T$ are the symbols at receiver's respective antennas which will be used to estimate transmitted symbols [3]. The receiver will analyse all received information to compute the transmitted data. So, a detection, computation, equalization and estimation of the received data will happen.

At receiver side, the linear zero forcing (ZF) detector compute the inverse of the channel matrix to estimate the transmitted symbols which can be expressed by,

$$\tilde{s}_{ZF} = \underbrace{(H^H.H)^{-1}.H^H}_{\text{Moore–Penrose pseudo–inverse}} .x \qquad (2)$$

The channel matrix $H$ is $QR$ decomposed into two parts as $H = QR$.

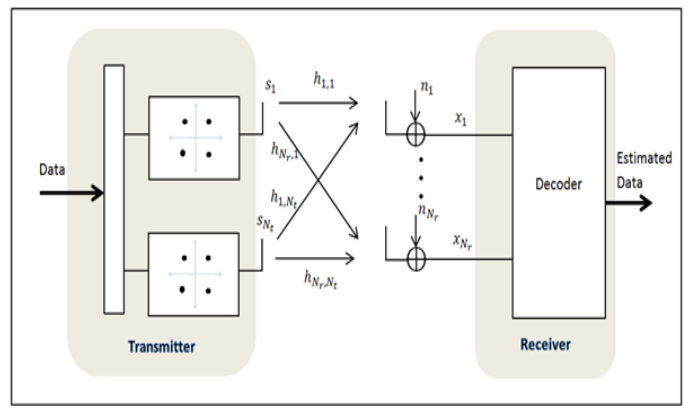

Figure 1. MIMO system with $N_t$ Transmitter and $N_r$ Receiver antennas.

## 3. LATTICE REDUCTION TECHNIQUE

We can interpret the columns $h_l$ of the channel matrix $H$ as the basis of a lattice and assume that the possible transmit vectors are given by $\mathbb{Z}^m$, the m dimensional infinite integer space. Consequently, the set of all possible undisturbed received signals is given by the lattice.

$$L(H) = L(h_1, \dots, h_m) := \sum_{l=1}^{m} h_l \mathbb{Z} \tag{3}$$

The LR algorithm generates a lattices reduced and near-orthogonal channel matrix $\widetilde{H} = H.T$. With matrix $\widetilde{H} = H.T$ generates the same lattice as $H$, if and only if the m × m matrix T is unimodular [2], i.e. $T$ contains only integer entries and $det(T) = \pm 1$:

$$L(\widetilde{H}) = L(H) \Leftrightarrow \widetilde{H} = HT \text{ and } T \text{ is unimodular} \tag{4}$$

Also,

$$\widetilde{H}.T^{-1} = H \tag{5}$$

We can find multiple bases that can be included in the space $L$, and the goal of the LR algorithm is to find a set of least correlated base with the shortest basis vectors [5]. Initially, an efficient (but supposed not optimal) way to determine a reduced basis was proposed by Lenstra, Lenstra and Lovàsz [1]. Where they defined (LLL-Reduced): A basis $\widetilde{H}$ with $QR$ decomposition $\widetilde{H} = \widetilde{Q}.\widetilde{R}$ is called LLL-reduced with parameter δ with $(1/4 < \delta \leq 1)$, if

$$|\widetilde{R}_{i,j}| \leq \tfrac{1}{2}.|\widetilde{R}_{i,i}| \text{ for } 1 \leq i < j \leq m \tag{6}$$

And

$$\delta|\widetilde{R}_{j-1,j-1}|^2 \leq |\widetilde{R}_{j,j}|^2 + |\widetilde{R}_{j-1,j}|^2 \text{ for } j = 2, \dots, m \tag{7}$$

The first condition is called, size-reduced and the second one is called Lovàsz condition. The parameter $\delta$ plays an important role to the quality of the reduced basis. We will assume $\delta = 3/4$ as proposed in [1]. After applying the $QR$ decomposition of $H$ and doing successive size-reduces operations if the condition is fulfilled, the algorithm exchanges two vectors if Lovàsz condition is not fulfilled to generate $T$ and compute $\widetilde{R}$ and $\widetilde{Q}$. And so, the LLL algorithm will output $\widetilde{Q}, \widetilde{R}$ and $T$.

Looking to the LLL algorithm [1], one important element of its complexity is related to the fact that the LLL algorithm is applied for the real integer vectors. It is mandatory to reformulate the different matrices to their real-valued form, so we got:

$$H^{real} = \begin{bmatrix} Real(H) & -Im(H) \\ Im(H) & real(H) \end{bmatrix}, \tag{8}$$

$$x = \begin{bmatrix} Real(x) \\ Im(x) \end{bmatrix}, \tag{9}$$

$$s = \begin{bmatrix} Real(s) \\ Im(s) \end{bmatrix} \text{ and } n = \begin{bmatrix} Real(n) \\ Im(n) \end{bmatrix} \tag{10}$$

International Journal of Wireless & Mobile Networks (IJWMN) Vol. 8, No. 3, June 2016

This kind of reformulation increases the number of operations and adds more latency for the system.

The idea behind LR-aided linear detection is to consider the equivalent system model and perform the nonlinear quantisation on it [8]. In fact, if we combine equations (1) and (5), we can get:

$$x = \widetilde{H}.\underbrace{T^{-1}.s}_{z} + n \qquad (11)$$

With $z = T^{-1}.s$ the equivalent model and in this case $\widetilde{H}$ will represent a better channel quality. And so, the detector can be represented with an equivalent model with better performance due to the less noise enhancement increased by $\widetilde{H}$. Thus, the basic idea behind approximate lattice decoding (LD) is to use LR in conjunction with traditional low-complexity decoders. With LR, the basis B is transformed into a new basis consisting of roughly orthogonal vectors [8].

After processing the Zero Forcing lattice reduction (ZF-LR) mechanism and by combining equations (2) and (11), we can generate:

$$\tilde{z}_{ZF-LR} = T^{-1}.\tilde{s}_{ZF} = \widetilde{H}.x = z + \widetilde{H}.n \qquad (12)$$

The different enhancements for the original algorithm were looking for a limited iterations in term of stopping criteria, like in [5]. But we believe that the structure of the triangular matrix generated by the QR decomposition can be an axe of improvement and complexity reduction.

Table 1. The LLL algorithm

| | |
|---|---|
| **Input:** | H: the channel matrix converted to the real-valued form |
| **Output:** | $\widetilde{R}, \widetilde{Q}, T$ |
| 1 | Initialization: $T = I_{N_t}$; Keeping in mind the real valued; H matrix |
| 2 | $[Q, R] := qr(H);$ |
| 3 | $\widetilde{R} = R; \widetilde{Q} = Q;$ |
| 4 | $\delta = 3/4$ |
| 5 | $k = 2$ |
| 6 | while $k \leq N_t$ |
| 7 | for $l = k - 1$ downe to 1 |
| 8 | $\mu := \lvert\widetilde{R}(l,k)/\widetilde{R}(l,k)\rvert$ |
| 9 | if $\mu \neq 0$ |
| 10 | $\widetilde{R}(1:l,k) := \widetilde{R}(1:l,k) - \mu.\widetilde{R}(1:l,l)$ |
| 11 | $T(:,k) := T(:,k) - \mu.T(:,l)$ |
| 12 | end |
| 13 | end |
| 14 | if $\delta.\widetilde{R}(k-1,k-1)^2 > \widetilde{R}(k,k)^2 + \widetilde{R}(k-1,k)^2$ |
| 15 | $k$ to $k-1$ columns swap for $\widetilde{R}$ and $T$ |
| | Computing the Givens rotaion Matrix: |
| 16 | $\Theta = \begin{bmatrix} \alpha & \beta \\ -\beta & \alpha \end{bmatrix}$ with $\alpha = \dfrac{\widetilde{R}(k-1,k-1)}{\lVert\widetilde{R}(k-1:k,k-1)\rVert}$, $\beta = \dfrac{\widetilde{R}(k,k-1)}{\lVert\widetilde{R}(k-1:k,k-1)\rVert}$ |
| 17 | $\widetilde{R}(k-1:k,k-1:m) := \Theta.\widetilde{R}(k-1:k,k-1:m)$ |

```
18            Q̃(:, k − 1: k) ≔ Q̃(:, k − 1: k). Θ^T
19            k ≔ max{k − 1, 2}
20         else
21            k ≔ k + 1
22         end
23      end
```

### 3.1. Exploiting R matrix's structure to improve the LLL algorithm

As shown in table 1 the outputs of the algorithm will be $\tilde{R}$, $\tilde{Q}$, $T$. With $\tilde{R}$ is an upper triangle matrix. The relation between them will follow (5).

$$\tilde{Q}.\tilde{R}.T^{-1} = Q.R = H \qquad (13)$$

Looking to the LLL algorithm, at lines 4 & 5 we can see that the loop is starting from $k = 2$. This choice is taken to reach the first column of $R$. This means that we can start from any other column $> 2$ and in this case we will not perform the column swap of columns $1$ $to$ $k - 2$.

So, in the case that the loop starts from 3; we will not perform column swap for first column. In this case we will gain 1 loop iteration and we will reduce the column swaps at least by 1. Looking to the morphology of the matrix $R$ which is a triangle matrix, so the first column contains only 1 active element (the rest are "0"). The major number of active elements is in the rest of the matrix.

### 3.2. R matrix's structure

Below, is a representation of the matrix $R$ in the case of $4 \times 4$ MIMO system. The number of elements by column is increasing from left to right

$$R = \begin{bmatrix} \overbrace{R_{1,1} \quad R_{2,1} \quad R_{3,1} \quad R_{4,1}}^{R^{1,4}} & \overbrace{R_{5,1} \quad R_{6,1} \quad R_{7,1} \quad R_{8,1}}^{R^{5,8}} \\ 0 & R_{2,2} & R_{3,2} & R_{4,2} & R_{5,2} & R_{6,2} & R_{7,2} & R_{8,2} \\ 0 & 0 & R_{3,3} & R_{4,3} & R_{5,3} & R_{6,3} & R_{7,3} & R_{8,3} \\ 0 & 0 & 0 & R_{4,4} & R_{5,4} & R_{6,4} & R_{7,4} & R_{8,4} \\ 0 & 0 & 0 & 0 & R_{5,5} & R_{6,5} & R_{7,5} & R_{8,5} \\ 0 & 0 & 0 & 0 & 0 & R_{6,6} & R_{7,6} & R_{8,6} \\ 0 & 0 & 0 & 0 & 0 & 0 & R_{7,7} & R_{8,7} \\ 0 & 0 & 0 & 0 & 0 & 0 & 0 & R_{8,8} \end{bmatrix} \qquad (14)$$

Let's decompose it schematically as 2 parts, $R^{1,4}$ and $R^{5,8}$ (Just to mention that the above choice is arbitrary).

In $R^{5,8}$, we have 26 active elements and in $R^{1,4}$ we have only 10 active elements. So, if we consider $R^{5,8}$ we can get 72% of the matrix elements. Adding the $R_{4,x}$ column we can get 30 elements and so 83% of the matrix elements. We are adding $R_{4,x}$ to be conforming to lines 6 to 11, 13 and 16 to 17 in table 1; if we consider $k = 5$.

If we consider a new matrix $R_{Limited}$ that conisists of the elements of R from column $R_{4,x}$ to $R_{8,x}$, so we will get a matrix $5 \times 8 R_{Limited}$. Which consist of 83% of actives elements of R. Thus, at the output of the LLL algorithm we will generate a $T_{Limited}$ matrix with 3 first elements as $I_3$ and $\tilde{R}$ will keep the firsts 3 elements of $R$.

$$\tilde{R} = \begin{bmatrix} R_{1,1} & R_{2,1} & R_{3,1} & \tilde{R}_{4,1} & \tilde{R}_{5,1} & \tilde{R}_{6,1} & \tilde{R}_{7,1} & \tilde{R}_{8,1} \\ 0 & R_{2,2} & R_{3,2} & \tilde{R}_{4,2} & \tilde{R}_{5,2} & \tilde{R}_{6,2} & \tilde{R}_{7,2} & \tilde{R}_{8,2} \\ 0 & 0 & R_{3,3} & \tilde{R}_{4,3} & \tilde{R}_{5,3} & \tilde{R}_{6,3} & \tilde{R}_{7,3} & \tilde{R}_{8,3} \\ 0 & 0 & 0 & \tilde{R}_{4,4} & \tilde{R}_{5,4} & \tilde{R}_{6,4} & \tilde{R}_{7,4} & \tilde{R}_{8,4} \\ 0 & 0 & 0 & 0 & \tilde{R}_{5,5} & \tilde{R}_{6,5} & \tilde{R}_{7,5} & \tilde{R}_{8,5} \\ 0 & 0 & 0 & 0 & 0 & \tilde{R}_{6,6} & \tilde{R}_{7,6} & \tilde{R}_{8,6} \\ 0 & 0 & 0 & 0 & 0 & 0 & \tilde{R}_{7,7} & \tilde{R}_{8,7} \\ 0 & 0 & 0 & 0 & 0 & 0 & 0 & \tilde{R}_{8,8} \end{bmatrix} \quad (15)$$

$$T = \begin{bmatrix} 1 & 0 & 0 & T_{4,1} & T_{5,1} & T_{6,1} & T_{7,1} & T_{8,1} \\ 0 & 1 & 0 & T_{4,2} & T_{5,2} & T_{6,2} & T_{7,2} & T_{8,2} \\ 0 & 0 & 1 & T_{4,3} & T_{5,3} & T_{6,3} & T_{7,3} & T_{8,3} \\ 0 & 0 & 0 & T_{4,4} & T_{5,4} & T_{6,4} & T_{7,4} & T_{8,4} \\ 0 & 0 & 0 & T_{4,5} & T_{5,5} & T_{6,5} & T_{7,5} & T_{8,5} \\ 0 & 0 & 0 & T_{4,6} & T_{5,6} & T_{6,6} & T_{7,6} & T_{8,6} \\ 0 & 0 & 0 & T_{4,7} & T_{5,7} & T_{6,7} & T_{7,7} & T_{8,7} \\ 0 & 0 & 0 & T_{4,8} & T_{5,8} & T_{6,8} & T_{7,8} & T_{8,8} \end{bmatrix} \quad (16)$$

This means that we have generated only 40 from 64 possible matrix element and only 5 from 8 possible columns for the matrix *T*. Consequently, for matrix *R* we have manipulated only 30 from 36 possible active elements. This is a considerable computation relaxation.

This approach can be generated for all column indexes which allow to gain more operations, and so we can change the algorithm of table 1 as below.

Table 2. The LLL algorithm with modified start point.

| | | |
|---|---|---|
| **Input:** | | *H*: the channel matrix converted to the real-valued form |
| **Output:** | | $\tilde{R}, \tilde{Q}, T$ |
| 1 | | Initialization: $T = I_{N_t}$; |
| 2 | | $[Q, R] := qr(H)$; |
| 3 | | $\tilde{R} = R; \tilde{Q} = Q$; |
| 4 | | $\delta = 3/4$ |
| 5 | | $k_{start} = (N_t/2)$ ; *or any value > 2* |
| 6 | | $k = k_{start}$ |
| 7 | | while $k \leq N_t$ |
| 8 | |   for $l = k - 1$ downe to 1 |
| 9 | |     $\mu := |\tilde{R}(l, k)/\tilde{R}(l, k)|$ |
| 10 | |     if $\mu \neq 0$ |

```
11                    R̃(1:l, k) := R̃(1:l, k) − μ . R̃(1:l, l)
12                    T(:, k) := T(:, k) − μ . T(:, l)
13                end
14            end
15            if δ. R̃(k − 1, k − 1)² > R̃(k, k)² + R̃(k − 1, k)²
16                k to k − 1 columns swap for R̃ and T
                  Computing the Givens rotaion Matrix:
17                Θ = [ α   β ]  with   α = R̃(k − 1, k − 1) / ‖R̃(k − 1: k, k − 1)‖
                      [−β   α ]         β = R̃(k, k − 1) / ‖R̃(k − 1: k, k − 1)‖
18                R̃(k − 1: k, k − 1: m) := Θ. R̃(k − 1: k, k − 1: m)
19                Q̃(:, k − 1: k) := Q̃(:, k − 1: k). Θᵀ
20                k := max{k − 1, k_start}
21            else
22                k := k + 1
23            end
24        end
```

But we should note that, logically the BER performance degradation will increase. In fact, we have some compromises to take into consideration (operations vs performance balance). Also, this approach will be more efficient as much as we use more antennas for both sides of the system. This means that we need to evaluate the cases where the approach will be beneficial in terms of complexity while keeping an acceptable performance.

In the next sections we will present the simulation results and the complexity study of the proposed approach.

## 4. SIMULATION RESULTS

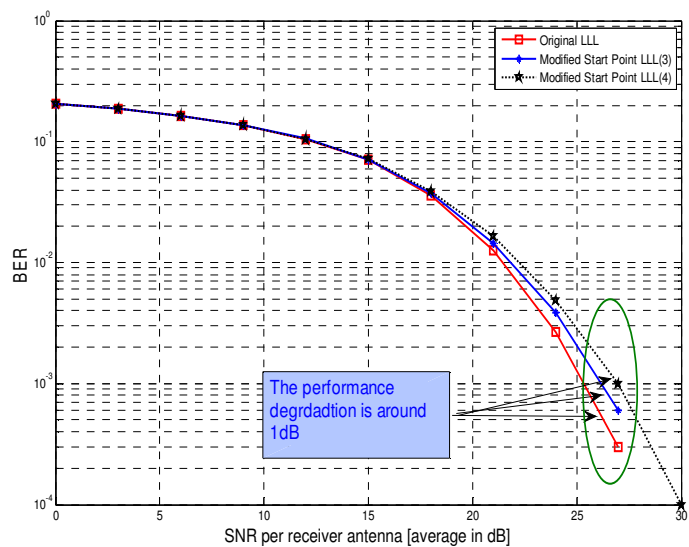

Figure 2. The BER performance comparison between the original LLL and the modified LLL algorithm ($k_{start} = 3$ and $k_{start} = 4$). Simulation related to 16QAM $4 \times 4$ MIMO system.

Figure 2, shows that if we consider the LLL algorithm starting point at column 3 or 4, the BER is not dramatically degrading (limited). But we gain a lot in terms of computation operations. In fact the proposed modification will avoid that the algorithm do more iterations and operations to reduce the elements of *R* (simultaneously to generate *T*), especially for the vectors without big effects on the results (performance). So, we will focus on the matrix column with maximum of active elements and avoiding making operation with almost "zeros" valued columns.

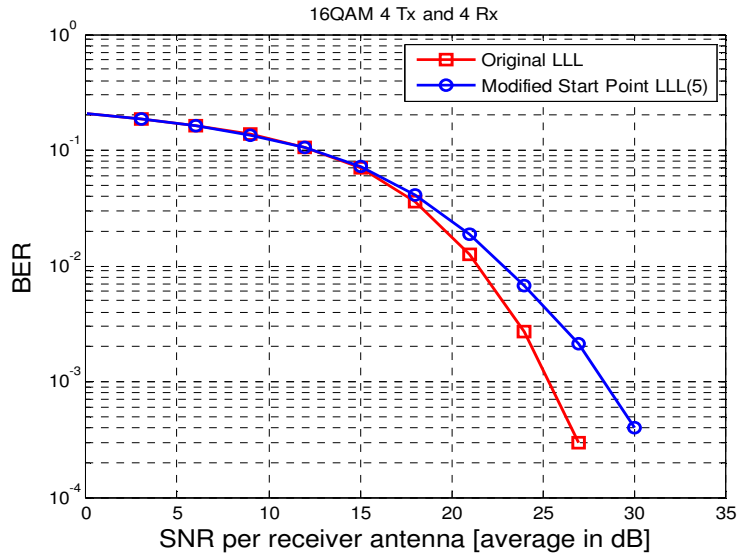

Figure 3. The BER performance comparison between the original LLL and the starting point modified LLL algorithm ($k_{start} = 5$). Simulation related to 16QAM $4 \times 4$ MIMO system.

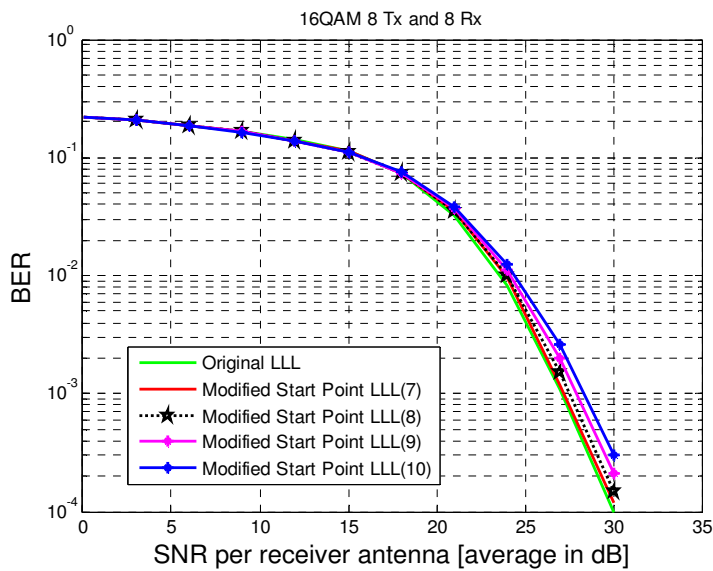

Figure 4. The BER performance comparison between the original LLL and the starting point modified LLL algorithm ($k_{start} = 7$ & $k_{start} = 9$). Simulation related to 16QAM $8 \times 8$ MIMO system.

Figure 4 shows the BER performance for the same approach applied to an 8×8 MIMO system. It illustrates clearly that the approach can be applied for any $N_t \times N_r$ MIMO system. Also, we can observe that for big sized matrixes the approach is showing better results. In fact, as much as we increase the matrix size we have more "zeroes" in the first columns and more "non-zeroes" elements for the right part of the matrix. So, we will get more possibilities to shift the start column index.

## 5. COMPLEXITY GAIN

In this section we will present an analysis of the operations load of the algorithm while being executed. After that, we will show the gain in term of operations and complexity that we can make after applying our proposed approach.

### 5.1 Operations analysis

By looking to the algorithm in table 1, we can observe that:

- The size reduction operations (lines 7 to 13), is doing a kind of loop with $k - 1$ iterations for a set of operation that contains; a division, two subtractions (than can be considered as addition [6]) and two multiplications if the line 9 is valid. So, in maximum of cases, the size reduction can be done with $N^{Reduction}$ operations.

$$N^{Reduction} = (k - 1) \times \{1 \times (Div) + 2 \times (Mult + Add)\} \quad (17)$$

- Line 14 representing Lovàsz's conditions require:

$$N^{Lovàsz} = \{4 \times (Mult) + 2 \times (Add)\} \quad (18)$$

Considering that the superior verification can be achieved via a subtractions operation [6].

- The columns permutation operation is being done elements by elements. Knowing that the simple two elements permutation is equivalent to three additions. Also, the algorithm doesn't make difference for zero or non-zero values. So, the column permutation will be done in:

$$N^{Permutation} = 2N_R \times \{3 \times (Add)\} \quad (19)$$

- The Givens rotation matrix corresponds to the computation of the $\alpha$ and $\beta$ parameters and this is being done via a "norm" calculation from one side, which corresponds to a square root operation, two multiplications and one additions. And, two divisions from another side.

$$N^{Givens} = \{2 \times (Div) + 2 \times (Mult) + 1 \times (Add) + 1 \times (sqrt)\} \quad (20)$$

- The $\tilde{R}_{Unit}$ unitary matrix and the rotation matrix multiplication correspond to a matrix $2 \times 2$ and matrix $2 \times (2N_R - k + 2)$ multiplications [5].

$$N^{Rotation} = 2 \times \{2 \times (Mult) + 1 \times (Add)\} \times (2N_R - k + 2) \quad (21)$$

- Line 18 corresponds to multiplication of two "$2 \times 2$" matrixes.

$$N^Q = 2 \times \{2 \times (Mult) + 1 \times (Add)\} \times 2 \qquad (22)$$

- Line 19 corresponds to a comparison and an assignment.

$$N^{max} = 2 \times (Add) \qquad (23)$$

This doesn't take in consideration the constellation, since we are using the same constellation for all the paper and so the analysis remains the same.

## 5.2 Flops analysis

Ameer and al [7] and Markus in [6] indicated a kind of correspondence between the operation and the number of flops.

Table 3. FLOPS vs operations

| Operation | Add | Mult | Sqrt | Div |
|---|---|---|---|---|
| Number of FLOPS | 1 | 1 | 8 | 8 |

The tables below will show the flops needed by a MIMO $4 \times 4$ and MIMO $8 \times 8$ systems and the gain that we can get after a start column shift.

Table 4. Gain after column shift for MIMO $4 \times 4$

|  | *No shift* | *shift $k = 3$* | *shift $k = 4$* |
|---|---|---|---|
| **MIMO $4 \times 4$** | 931 | 822 | 715 |
| **Gain in %** | 0% | 11.7% | 23% |

Table 5: Gain after column shift for MIMO $8 \times 8$

|  | *No shift* | *shift $k = 3$* | *shift $k = 4$* |
|---|---|---|---|
| **MIMO $8 \times 8$** | 3075 | 2430 | 2090 |
| **Gain in %** | 0% | 21% | 32% |

As mentioned above, in the case of an upper triangular matrix, almost of the first columns elements are "zeros". Also, the columns permutation and matrix multiplication in the algorithm don't make a difference for "zero" and "non-zero" elements. So, it makes a lot of additions of element with zero or a multiplication by zero, etc.… Thus, the operation done in the first columns will consume a lot of resources while its income in terms of information is limited.

With our approach we target to avoid the non-useful operations (first columns which are full of "zeros") and concentrate the effort on the columns with the maximum of information. Our approach shows a good operations gain (which equivalent to complexity in this case) and good performances (the BER results).

In the case of MIMO $8 \times 8$ system we can reduce 32% of operations, when doing a column shift of 9 columns, with a very limited BER degradation (less than 2dB).

We should note that in this paper, we didn't present the case of MIMO $2 \times 2$ system. This was related to the size of the matrix which is small and the new algorithm will not bring a considerable outcome. Keeping in mind that the LLL algorithm is mainly used to simplify the decoding with "big size" channels matrixes [2].

## 6. CONCLUSION

In this paper, we proposed a modified LLL algorithm that exploits a kind of shift start column. We started from the original LLL algorithm and we modified it to escape the almost "zeros" columns of the upper triangular matrix R. And so, we avoided doing computation for the columns without big influence on the BER performance. The proposed approach is not one of the fashion modification of the LLL algorithm, but its added value come from its simplicity and complexity gain.This approach was simulated for both 4×4 and 8×8 MIMO systems and can be extended for any other MIMO system model. We have presented that we can gain, respectively, 23% and 32% for the MIMO 4×4 and MIMO 8×8 scheme. This is an important point, we are reducing the computation operations and so the decoding time with very limited BER degradation. So, the trade-off complexity vs performance is interesting and the gain in terms of complexity counterbalance the limited performance degradation.

We considered the 16QAM modulation and ZF receiver where our approach shows good results. It will be interesting to extend this study to the MMSE and other modulation techniques. Also, in this paper, we discussed the case with a same antennas number on both sides. The case with a different antennas number on both sides will be the subject of a new study. Finally, this approach is showing better results for the "big size" MIMO systems and it we believe that extend it to the case of massive MIMO will be interesting.